\documentclass[
    ,final            
  ]
  {aipproc}

\layoutstyle{6x9}

\begin{document}

\title{Do halos exist on the dripline of deformed nuclei?}

\classification{21.10.-k,21.10.Dr,21.45.+v,21.60.-n}
\keywords      {Halo Nuclei, three-body models, HFB, deformation, pairing }

\author{F.M. Nunes, B. Avez, T. Duguet}{
  address={National Superconducting Cyclotron Laboratory and 
  Department of Physics and Astronomy, \\
  Michigan State University, East-Lansing MI 48824-1321}
}

\begin{abstract}
A study of the effect of deformation and pairing on
the development of halo nuclei is presented.
Exploratory three-body $core+n+n$ calculations show that
both the NN interaction
and the deformation/excitation of the core hinder the 
formation of the halo. Preliminary self-consistent mean-field calculations
are used to search for regions in the nuclear chart where
halos could potentially develop. These are also briefly discussed.
\end{abstract}

\maketitle


\section{Motivation}

During the early years of Radioactive Nuclear Beam physics, 
while the dripline for light nuclei was being explored, 
{\it nuclear halos} became a very hot topic \cite{halo}.
Typical examples of halo nuclei include $^6$He, $^{11}$Li, $^{14}$Be
and $^{19}$C, all in the low mass region of the nuclear chart.
The big open question,
when moving toward heavier systems,  is whether
halos can develop when for instance $A>40$.

In order to successfully describe a halo nucleus, the structure model
needs to take into account \cite{nunes03}: 
i) the very low density region in which
the halo nucleons move, subject to an interaction that is closer to the
free NN interaction than the realistic in-medium nuclear interaction; 
ii) the long tails of the wavefunctions and correct asymptotics
of these tails, which contribute decisively to many nuclear properties;
iii) the few-body dynamics of the valence nucleons relative to
the core and between themselves.
For these reasons,  it is acceptable to decouple the halo degrees of 
freedom from the core's, simplifying the standard microscopic treatment: 
this is the basis for applying few-body models to light nuclear halos.

In heavier systems, the decoupling of core and valence nucleons
may not be as straightforward. Mean-field studies
in the past decades have shown that
pairing is important, and in some
regions of the nuclear chart (namely the deformed region), 
deformation is also necessary.
As matter densities in Hartree Fock Bogolyubov (HFB) have a faster
radial decay when pairing is included, one expects 
the pairing force to act against the halo formation.
The pairing anti-halo effect was shown in Ref. \cite{benna}
through self-consistent Hartree Fock Bogolyubov calculations
for Carbon isotopes.
In opposition to \cite{benna}, the results of Hamamoto et al. \cite{hama1,hama2} 
indicate that when an $s_{1/2}$ neutron  approaches the dripline, 
it becomes decoupled from the mean field, which means they are less paired
and consequently may give rise to a halo. One can then claim
that, if the neutron becomes decoupled from the mean field,
the HFB theory should not be used. Unfortunately,
the calculations in \cite{hama1,hama2} are not self-consistent;
the mean field is reduced arbitrarily, from a well bounded
situation, to force the system toward the dripline.

If the situation with pairing is controversial, the effect
of deformation appears to be well settled.
Nilsson model calculations performed by Hamamoto \cite{hama3} 
suggest that  \mbox{$J_z=1/2^+$} states 
($J_z$ being the projection of the angular momentum on the
deformation axis), which contain s-wave components
of the valence particle as well as other components with higher
partial waves,  become pure s-waves when the system it
forced toward threshold. Thus, it is concluded  that
deformation does not hinder the halo formation \cite{hama3}.

This exotic feature is not solely associated to nuclear physics:
halo manifestations appear also in atomic, molecular or condensed matter 
physics. A detailed review \cite{jensen}
compiles the results across the nuclear border, and develops
a halo signature that removes the specific scale (nuclear, atomic, etc).

\section{Results with three-body Model}

In Ref. \cite{nunes05} we study the structure of the ground state of a nucleus 
with two valence nucleons as the system approaches the two particle
threshold. A three-body model of $core+n+n$ is used
where the core is deformed and allowed to excite.  The Faddeev Equations
are solved within the hyperspherical method. 
Our starting point is the $^{12}$Be model \cite{nunes96}.
An important ingredient of the model \cite{nunes96} 
is the core+n interaction fitted to reproduce the properties of $^{11}$Be.
We artificially decrease this $core-n$ interaction,
\begin{equation}
V_{n-core}(\vec r) = \lambda \; V^{be12}_{n-core}(\vec r)\;.
\end{equation}
with $\lambda \rightarrow 0$ to simulate the approach to
the neutron dripline \cite{nunes05}.

The correspondence from a few-body language to the mean-field
terminology appropriate for heavy systems, is far from trivial.
Nevertheless, in some way, pairing and deformation are included 
in this three-body model. In the limit of a very heavy core, the core-n
interaction plays the role of the mean field in the microscopic
description. The NN interaction $V_{nn}$ included in the three-body model
would then be related to the pairing associated with 
the valence pair only, in the mean-field language. However 
the pairing force and $V_{nn}$ are not identical. Here $V_{nn}$ is
a sum of gaussians fitted to the low energy NN phases shifts,
containing central s, p and d terms, as well as a spin orbit
and a tensor force, while pairing in a typical mean-field
calculation is zero range and s-wave only
(see \cite{nunes05,duguet04} for a further discussion).
\begin{figure}[t!]
  \includegraphics[height=.3\textheight]{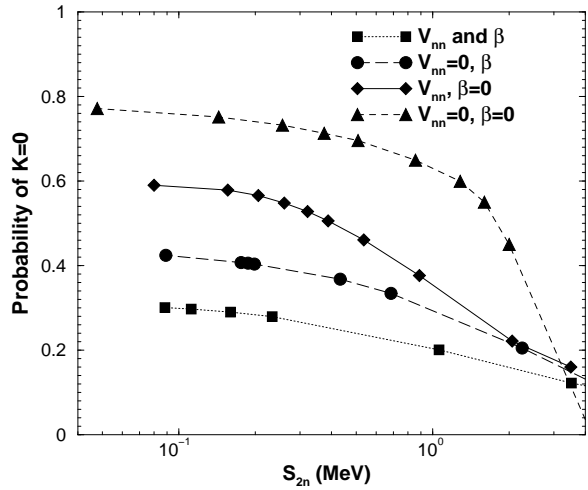}
  \caption{Probabilities for the K=0 component in the 
  ground state wavefunction
  of a $^{12}$Be-like system as a function of the two valence
  neutron separation energy.}
\end{figure}

The hyperspherical expansion introduces a new quantum number
$K$ related to the sum of the angular momenta associated
with the Jacobi coordinates (x,y) \cite{nunes96}: $K=(l_x+l_y)/2$. 
The resulting coupled channel equations
contain a centrifugal barrier of the form $(K+3/2)(K+5/2)/\rho^2$,
where $\rho=\sqrt{x^2+y^2}$ is the size of the halo.
Even for the lowest hyper momentum $K=0$, there is
a barrier reducing the halo effect.
For a halo to appear,
the K=0 component needs to be the dominant component (>50\%) \cite{jensen}. 
Thus, we look at the
probability of K=0 in the ground state wavefunction as a function
of the three-body binding energy (Fig. 1). 
The probabilities for the lowest hyperspherical
harmonic component (K=0) as a function of the two-neutron separation
energy are shown for four different cases:
i)
the case where a realistic NN interaction is included, as well
as a quadrupole deformation $\beta_2=0.67$ for the $^{10}$Be
core that reproduces the experimental $B(E2; 0^+ \rightarrow 2^+)$; 
ii)
the case where only the NN interaction is switched off;
iii)
the case where deformation/excitation is neglected 
but the NN interaction is included;
iv)
the case where both the NN interaction and core deformation/excitation
are neglected.

The first thing that should be noted is that, even for no
deformation/excitation and for no NN interaction, the system
never develops a pure K=0 state in the limit of zero binding.
This result is associated with the fact that we are treating
the system as a three-body system and it is in contrast 
with what would be obtained in the two body case \cite{jensen}. 
Secondly,  both the NN interaction and 
collective core degrees of freedom decrease considerably 
the limiting value for Prob(K=0) when $S_{2n}\rightarrow0$. 
Ultimately, when both effects are included, 
the probability is well below $50$ \%, suggesting that no halo will appear.

Note that the physical input for this exploratory study was provided by
$^{12}$Be. One can repeat the calculations reducing the deformation
parameter and increasing the mass of the core, to simulate
a heavier system more realistically \cite{nunes05}. 
However the model would still be inconsistent, in the sense
that the core-n interaction would not be determined from 
the physical subsystem, and the approach to the dripline 
would remain artificial.

\section{Preliminary HFB results}

The appropriate way to address the problem for medium mass to 
heavy nuclei is through a self-consistent
model that contains both pairing and deformation \cite{bender}. 
Let us first concentrate on pairing only.
A systematic HFB study was performed, searching for nuclei 
with the $3s_{1/2}$ neutron orbital
intercepting the Fermi level in the limit of stability \cite{benoit}. 
All calculations were performed with 
the HFB code written in a 3D mesh \cite{hfbcode}.
We found that the Cr isotopes satisfied this condition. 
Using SLy4 \cite{skyrme} for the particle-hole channel and 
ULB zero-range density-dependent pairing \cite{ulb} 
for the particle-particle channel,
the neutron dripline is reached for $^{80}$Cr. 
We plot the proton and neutron 
Helm radii \cite{helm} and geometrical radii (Fig. 2)
as a function of isotopic mass. 
The large difference between the neutron and proton Helm radii
indicate a neutron skin. A neutron halo exists when there is
a large difference between the neutron Helm  and
geometrical radii. 
We see that both are present on the Cr dripline.
A detailed analysis of our results do not confirm those from 
Ref. \cite{hama1} concerning the decoupling of the s-wave orbital 
from the core, when reaching the dripline \cite{benoit}.
\vspace{0.5cm}
\begin{figure}[h!]
  \includegraphics[height=.3\textheight]{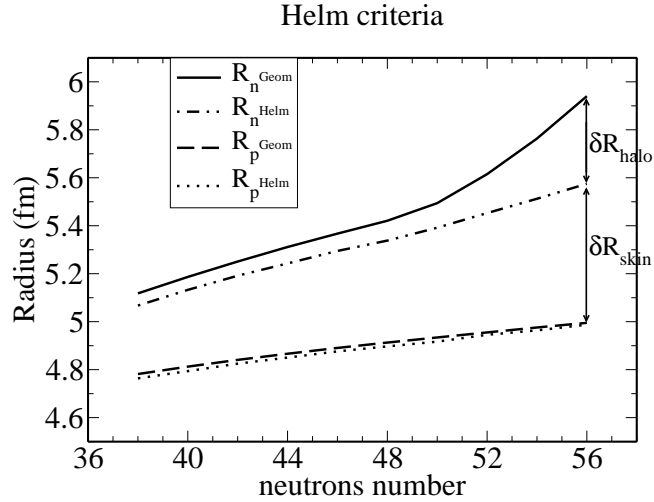}
  \caption{Helm criteria for halos and skins: geometrical and
  Helm radii for both neutrons and protons in Cr isotopes.}
\end{figure}
\begin{figure}[h!]
  \includegraphics[height=.3\textheight]{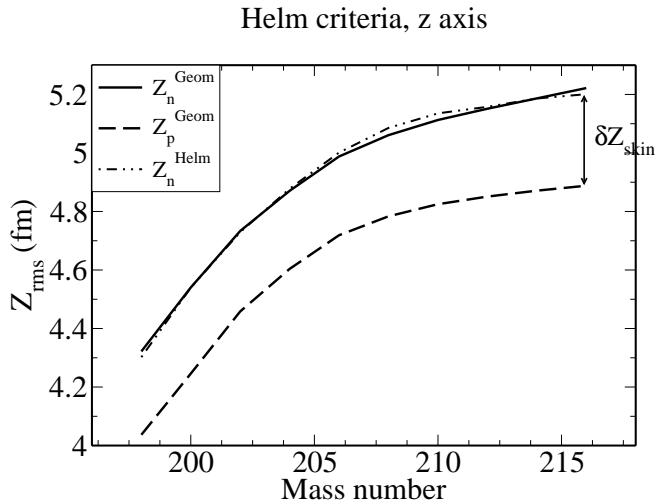}
  \caption{Same as Fig.(2) for Dy isotopes along the deformation axis.}
\end{figure}

Similar searches where performed in the region of deformed
nuclei using HFB with even multipole deformation \cite{hfbdef}.
For the deformed region the orbital we need to concentrate on
is the harmonic oscillator $1/2^+$ Nilsson orbital coming 
from the $4s_{1/2}$ shell
in the limit of sphericity. Only for the prolate Dy isotopes 
did we find this valence neutron orbital to be close
to the Fermi level at the dripline. The analysis  based on the Helm
and geometrical radii was performed on the Dy isotopes (see Fig. 3),
similarly to the Cr case. For that purpose, an extension of the Helm
criterion to deformed systems was proposed \cite{benoit}.
As there is no significant difference between the neutron's
Helm and geometrical radii, we expect no halo to develop.
Note that Dy isotopes demonstrate another minimum for oblate shape,
which means that the correct ground state should have configuration
admixture. Such a calculation was not performed due to its complexity,
yet we do not expect it would reverse the result concerning halos.

From these preliminary self-consistent HFB studies we can draw the following
conclusions:
\begin{itemize}
\item
In a realistic situation, pairing does indeed oppose the
development of the halo. This confirms the results of \cite{benna}
and the three-body model \cite{nunes05}. The effect is not 
sufficiently strong to hinder completely a halo in $^{80}$Cr.
\item 
The self-consistent HFB results do not show the decoupling of the 
neutron from the mean field seen by \cite{hama1}. 
\item
Deformation in heavy nuclei seem to strongly hinder
the appearance of a halo, as suggested by the 
three-body model \cite{nunes05} and in disagreement with
\cite{hama3}. No halos were
found in the deformed region of the nuclear chart.
In particular, the fully self-consistent calculations
do not confirm the purification of the $J_z=1/2^+$ orbitals at the
Fermi level into s-waves when approaching the dripline.

\end{itemize}
An article containing all details of these HFB studies will become
available soon. As part of our near future plans, we will study
the dependence on the choice of the microscopic pairing force,
namely a better low-density behaviour and finite-range effect
\cite{duguet04}.

\vspace{0.5cm}
\small
$^{*}$ This work has been partially
supported by National Superconducting Cyclotron Laboratory at 
Michigan State University and the National Science Foundation 
under grants PHY-0456656 and PHY-0456903.

\end{document}